**Domain Wall Roughness and Creep in Nanoscale Crystalline Ferroelectric Polymers**

Z. Xiao[1], Shashi Poddar[1], Stephen Ducharme[1,2], and X. Hong[1,2*]

[1] Department of Physics and Astronomy, University of Nebraska-Lincoln, NE 68588-0299

[2] Nebraska Center for Materials and Nanoscience, University of Nebraska-Lincoln, NE 68588-0299

**Abstract:**

We report piezo-response force microscopy studies of the static and dynamic properties of domain walls (DWs) in 11 to 36 nm thick films of crystalline ferroelectric poly(vinylidene-fluoride-trifluorethylene). The DW roughness exponent $\zeta$ ranges from 0.39 to 0.48 and the DW creep exponent $\mu$ varies from 0.20 to 0.28, revealing an unexpected effective dimensionality of ~1.5 that is independent of film thickness. Our results suggest predominantly 2D ferroelectricity in the layered polymer and we attribute the fractal dimensionality to DW deroughening due to the correlations between the in-plane and out-of-plane polarization, an effect that can be exploited to achieve high lateral domain density for developing nanoscale ferroelectrics-based applications.

[*] Xia Hong: xhong2@unl.edu

The finite size effect related to the critical thickness for ferroelectric instability has been a central topic of debate in modern ferroelectric studies, and sets the fundamental scaling limit on ferroelectric-based nanoelectronic devices such as ferroelectric tunnel junctions.[1,2] New ground states and domain structures have been predicted and observed in ultrathin perovskite ferroelectrics such as PbTiO$_3$ and BaTiO$_3$, and the results depend highly on the experimental details of sample growth and characterization.[2] Another class of ferroelectrics, polymers such as poly(vinylidene–fluoride–trifluorethylene) (PVDF-TrFE), presents a more robust system for investigating size scaling, as the local breaking of inversion symmetry is inherent to the constituent molecular structure. PVDF-TrFE are random copolymers consisting of long chains of the form $-((-CF_2-CH_2)_x-(-CF_2-CHF-)_{1-x})_n-$, where polarization depends on the orientations of $-CF_2-CH_2-$ dipoles (Fig. 1(a)). These polymers can be prepared in layer by layer crystalline structures, and it is conceivable that they are immune to the finite size effect and can preserve ferroelectricity as the system approaches the two-dimensional (2D) limit.[3,4]

Critical information on how ferroelectricity evolves with the system dimension can be gained by studying the static configuration and dynamic response of ferroelectric domain walls (DWs). These properties of DWs also determine the fundamental density limit and ultimate operation speed of 2D ferroelectric devices. It has been shown that for a $d$-dimensional system, DWs can be treated as ($d$-1)-dimensional elastic manifolds that wander in the landscape of random disorder potential.[5,6,7] The static roughness of the DWs can be described by scaling behavior with a characteristic roughness exponent $\zeta$.[5] When subject to a small driving force $f$, the propagation of the DWs follows the nonlinear creep behavior with the velocity given by $v \propto \exp[-\frac{\Delta}{k_BT}(\frac{f_c}{f})^\mu]$, where $\Delta$ is a scaling energy constant and $f_c$ is the critical depinning force[5]. The DW roughness exponent $\zeta$ and creep exponent $\mu$ can reveal information on the dimensionality and dominating disorder of the system.[5,6,7]

In previous studies, direct imaging of DWs using optical or scanning probe approaches have been intensively investigated in magnetic systems[8,9,10] and ferroelectric and multiferroic oxides.[11-15] However, only few scanning probe studies have been carried out on polymeric ferroelectric thin films.[16,17,18,19] Questions such as the critical length and time scales for domain nucleation and propagation in crystalline polymer thin films as the system approaches lower dimensions, especially in the presence of disorder, remain to be answered.

In this letter, we report a nanoscale scanning probe study of DW roughness and creep behavior in polycrystalline PVDF-TrFE films prepared by Langmuir Blodgett (LB) approach. We have employed atomic force microscopy (AFM) and piezo-response force microscopy (PFM) to write and image ferroelectric DWs in 6-20 monolayer (ML) thick PVDF-TrFE films. The extracted DW roughness exponent $\zeta$ varies from 0.39 to 0.48. The creep exponent $\mu$ is close to 0.25, the predicted value for a 1D DW.[5] The critical exponents reveal an unexpected, thickness-independent effective dimensionality of $d_{\text{eff}} \approx 1.5$, which is in sharp contrast to the 1D and 2D DWs observed in ferroelectric oxide Pb(Zr,Ti)O$_3$ (PZT) films of similar thickness (30-100 nm).[11,12,14] The fractal dimensionality and weak dependence on the film thickness suggest that the interlayer interaction plays a minor role in ferroelectric domain nucleation and propagation, and we propose that the DW can be deroughened by the disordered in-plane component of the polarization. This type of correlated disorder is introduced by the intrinsic orientation of polarization, which can be easily implemented into crystalline ferroelectrics. Our results thus suggest an effective and relatively low cost route to achieve higher lateral density in nanoscale ferroelectric-based data storage and sensing devices.[20]

Figure 1(b) shows the schematic experimental setup. We evaporated 5 nm Cr and 50 nm Au on top of a $SiO_2$-Si wafer as the bottom electrode, and then deposited PVDF-TrFE film one nominal ML at a time using the LB technique.[3,4] The molar content ratio of 75:25 (PVDF:TrFE) was used to achieve high piezoelectric response. For this composition the ferroelectric Curie temperature $T_C$ is ~110°C. The samples were then annealed at 135°C in a forced air oven for 90 min following the procedure described in Refs. [4,16] to achieve a well-ordered structure. After annealing, the samples were polycrystalline within each layer with grain sizes on the order of 50 nm,[16] and each ML is approximately 1.8 nm thick.[21] For all grains, PVDF-TrFE chains are close-packed in an orthorhombic structure with the polar direction pointing 30° with respect to the surface normal.[4] AFM measurements show smooth surfaces with RMS roughness of ~1 nm for the 10 ML films (Fig. 1(e) inset). The DW studies are carried out using a Bruker Multimode 8 AFM. We use a low spring constant AFM probe (Bruker SCM-PIC, 0.2 N/m), which allows us to write and image ferroelectric domains without causing mechanical damage to the polymer. For imaging, we scan close to one of the resonant frequencies of the cantilever (170±20 kHz) with 1 V excitation voltage.

The as prepared PVDF-TrFE films have the out-of-plane polarization uniformly polarized to the up position before poling (Fig. 1(b)), while the in-plane polarization in different grains varies in orientation.[18] To study the static configuration of the DWs, we wrote stripe domains by scanning the film while applying voltage higher than the coercive voltage (<1 V/ML) with alternating polarities to the conducting tip. Figures 1(d) and 1(e) show the phase and amplitude images of oppositely polarized stripe domains on a 10 ML film. For 5 μm x 5 μm areas we image with 512 lines and 512 points/line (sampling interval of ~10 nm).[22] We extract the DW position (Fig. 1(f)) using both the middle of the signal levels of the phase response (Fig. 1(g)) and the lowest amplitude response (Fig. 1(h)).

We first examined the geometric fluctuation of the DWs by calculating the correlation function: $B(L) = \overline{\langle [u(L) - u(0)]^2 \rangle}$. Here $u(L)$ is the perpendicular displacement of the DW at position $L$ from the its flat configuration (Fig. 1(c)), and $B(L)$ is averaged over the DW longitudinal coordinates ($\langle \cdots \rangle$) and disorder ($\overline{\cdots}$). As shown in Fig. 2, $B(L)$ increases rapidly with $L$ at short length scales and then tends toward saturation at around $L = 50$ nm. This saturation behavior is because the PVDF-TrFE films are polycrystalline with an average grain size of 50 nm,[16] and we do not expect DWs among different crystalline grains to be correlated.

The rapid growth of $B(L)$ at the short length scale can be well described by a power law dependence. It has been shown that DWs can be considered as elastic with the correlation function described by $B(L) \propto L^{2\zeta}$ at a length scale larger than the characteristic Larkin length, which is on the order of the DW width or the relevant length of the pinning potential.[5,6] For a 15 ML film, we extracted the average roughness exponent of multiple DWs to be $\zeta = 0.42 \pm 0.03$ (Fig. 2(b)). The results obtained from the phase and amplitude images show excellent agreement with each other, confirming the qualitative behavior of the DW. For consistent comparison we extracted the $\zeta$ values in all samples based on data taken between $L = 10$ nm and 50 nm. However, we also note that for some samples the saturation of $B(L)$ actually starts below $L = 50$ nm (Fig. 2(b)), likely because we are sampling over grains with smaller sizes.[22]

The observed roughness exponent is significantly different from the values observed in 1D and 2D DWs in PZT films well below $T_C$.[12,14] Two universality classes of quenched disorders can play major roles in DW roughening.[12] Random bond disorder is short-ranged and modifies the ferroelectric double well energy symmetrically. The corresponding exponents are:

$$\zeta_{RB} = 2/3 \text{ for } d_{\text{eff}} = 1 \text{ and } \zeta_{RB} = 0.2084(4 - d_{eff}) \text{ for } d_{\text{eff}} > 1 \qquad (1a),$$

where $d_{\text{eff}}$ is the effective dimensionality of the system. Random field disorder makes the double well potential energy asymmetric and is effectively long-ranged,

$$\zeta_{RF} = (4 - d_{eff})/3 \qquad \text{for all dimensions} \qquad (1b).$$

The roughness exponent in PVDF-TrFE suggests that it is intrinsically different from PZT in either $d_{\text{eff}}$ or the type of disorder.[12,14]

To identify the origin of this difference, we have investigated how DWs propagate under an external electric field. We wrote dot-shaped domain structures on a uniformly polarized background (written with 11 V) on the 15 ML film by applying -11 V voltage pulses with different pulse durations (500 μs to 8 s) to a standing AFM tip. As shown in Fig 3(a), the resulting domain size increases with increasing pulse duration. We first obtained the radius $r$ of the dot-domains by calculating the domain area $S$ from the PFM phase image and using $r = \sqrt{S/\pi}$. In this way we can average out the non-circular effect of the domain shape, which is likely due to the polycrystalline grains. For domains created by two slightly different pulse durations $t_1$ and $t_2$, we extracted the corresponding domain radii $r_1$ and $r_2$, and approximated the transverse DW velocity using $v(r_0) = (r_2 - r_1)/(t_2 - t_1)$, where $r_0 = (r_1 + r_2)/2$. We then calculated the electric field $E$ at position $r_0$ using $E = Va/r_0d$, where $V$ is the AFM tip bias voltage, $a$ is the tip radius, and $d$ is the polymer film thickness. Here we assumed that the electric potential due to the biased AFM tip is spherical, which has been shown in previous studies to be a good approximation.[11] We found the DW velocity spans three orders of magnitude from $10^{-8}$ m/s to $10^{-5}$ m/s for electric fields ranging from $2\times10^7$ V/m to $1\times10^8$ V/m (Fig. 3(b)). As the radii of the dot-domains vary from ~40 nm to ~450nm, each DW velocity value is extracted based on averaging over a large number of grains.

It has been shown that when the driving electric field $E$ is well below the depinning electric field $E_0$, the velocity of the ferroelectric DW follows the non-linear creep model as $\propto \exp\{-\frac{\Delta}{k_B T}(E_0/E)^\mu\}$.[5,11] For the 15 ML film, the DW velocity is well described by the creep motion with a critical exponent $\mu = 0.21\pm0.03$. This value is consistent with the value $\mu = 0.23$ reported in a previous study on PVDF-TrFE nanomesas,[19] and close to the predicted value $\mu = 1/4$ for a 1D DW.[23] From the relation between the creep and roughness exponents: $\mu = (d_{eff} - 2 + 2\zeta)/(2 - \zeta)$,[5] we extracted the effective dimensionality of the system to be $d_{eff} = 1.5\pm0.2$. We then compared the experimental results with the $\zeta$ values obtained from the models of random bond disorder and random field disorder for $d = 1.5$ (Eq. (1)). As shown in Fig. 2(b), while the random bond disorder model provides a better description of our data, especially at the low $L$ regime, the value of $\zeta_{RB}$ is about 20% higher than the experimentally extracted value of $\zeta$. A possible origin for such discrepancy is related to our *a priori* choice of $L$-range (10 – 50 nm) for extracting $\zeta$, as the distribution of different polycrystalline grain sizes can cause early softening of $B(L)$ for $L < 50$ nm and the corresponding $\zeta$ just gives a lower bound of the exponent.[22] Another possibility is that the strength of the pinning potential is strong compared with the elastic energy of the system, and the roughness exponent derived for the weak disorder limit needs to be modified.[5] Further theoretical studies and experiments on PVDF-TrFE films with larger grain sizes are needed to clarify the nature of disorder in this system.

The extracted effective dimensionality is in sharp contrast with $d_{eff} = 2.5$ observed in DWs created and imaged under similar conditions in PZT films,[11,12,14] which corresponds to a 2D DW with an additional dimensionality of $(d-1)/2 = 0.5$ introduced by the long-ranged dipole interaction.[6] The creep exponent $\mu$ shows excellent agreement with the predicted value for a 1D disordered elastic system, suggesting that the ferroelectric polymer where DW propagates is essentially two-dimensional. However, the observed roughness exponent $\zeta = 0.42+/-0.03$ is clearly lower than the expected value of 2/3 for a 1D DW, and contributes to the fractal dimensionality of $d_{eff} = 1.5$. We also note that the dipole interaction does not change $d_{eff}$ for $d = 1$ DW systems.

To clarify the role of film thickness in this unusual $d_{eff}$, we have studied DWs in 6 to 20 MLs PVDF-TrFE films (11-36 nm). Experiments on thinner films are not available as films spontaneously break into nanomesa structures with discontinuous layer coverage.[24] As shown in Figs. 4(a) and 4(b), the roughness exponent $\zeta$ varies from 0.39 to 0.48 and the creep exponent $\mu$ ranges from 0.20 to 0.28. Both exponents do not show apparent dependence on the film thickness. The corresponding effective dimensionality ranges from 1.5 to 1.7 (Fig. 4(c)), very close to $d_{eff} = 1 + 1/2$. For each film thickness, we imaged 5 to 12 strip-shaped DWs in different locations on multiple samples, and examined 6 to 12 dot-shaped DWs for each pulse duration. The robustness of the exponent values indicates that sample preparation and imaging conditions do not significantly affect the observed scaling behavior, and the fractal $d_{eff}$ values are intrinsic to PVDF-TrFE rather than a transient effect due to film thickness scaling.

A likely mechanism that gives rise to the low dimensionality and weak thickness dependence is the strong anisotropy between the in-plane and inter-layer interactions. PVDF-TrFE LB films are polycrystalline within each monolayer. Within a crystalline grain, the polymer chains are closely packed in an orthorhombic structure, and have to switch collectively. The domain nucleation and propagation are thus dominated by the short-ranged elastic energy associated with the collective rotation of the polymer chains during the polarization switching.[25] However, two neighboring MLs are coupled with van de Waal force, and the inter-layer energy is dominated by the electrostatic energy due to the electric dipole interaction. For the current polarization orientation, the dipole interaction energy is on the order of 60 meV,[26] much smaller than the elastic energy of order 800 meV for twisting the polymer chain.[27] As a result, instead of the 3D ferroelectricity, PVDF-TrFE can be treated essentially as multi-layers of 2D ferroelectric systems that are weakly coupled to each other.[3]

We then consider what contributes to the unusually low roughness exponent that leads to the fractal dimensionality. As shown in Eq. (1), $\zeta$ carries the key information of the disorder type in the system. In previous studies on 2D magnetic systems, it has been observed that linear correlated disorder, which modifies the DW elasticity anisotropically, can deroughen the DWs [9,10] while keeping the dynamic exponent unchanged.[9] In PVDF-TrFE LB films, a special type of correlated disorder that can affect the elasticity of the system is introduced by the orientation of the molecular dipole. As the polarization is pointing 30° away from the surface normal, only the perpendicular component is aligned. During the polarization reversal, the in-plane dipoles have to rotate collectively along with the out-of-plane polarization and the final dipole orientation has to conform to the symmetry constraint imposed by the interaction between the substrates and the polymer.[4] Such constraint can be satisfied by both uncharged 180° DWs and charged 120° DWs. In addition, the polarization reversal can occur either through inter-chain motion or intra-chain twisting, which corresponds to either zigzag-shaped DWs or straight DWs, respectively.[27,28] The co-existence of different DW angles and configurations lead to anisotropic elasticity in the system, and can effectively lower DW roughness at the large scale. We propose that it is the correlation between the disordered in-plane component of the polarization and the out-of-plane polarization within each ML that lowers the roughness exponent of the 1D DW and contributes to the additional 0.5 in dimensionality. Interestingly, a similar value for DW $\zeta$ has been previously observed in multiferroic $BiFeO_3$ films (70 nm), and the role played by the magnetic order is not well understood.[15] Experiments that may clarify this effect include creating DWs with well-defined angles or introducing other types of correlated disorder into the system, and examining the corresponding critical exponents.

In conclusion, we have studied the DW roughness and creep behavior in 6-20 ML thick ferroelectric polymer films. The roughness and creep exponents of the DWs suggest that the polymers show predominantly 2D polarization. We also observe a fractal dimensionality of 1.5, which is attributed to the correlation between the in-plane and out-of-plane polarization. As the DW width in these polymers is at the sub-nanometer scale,[27,28] we expect that the domains can be stabilized at sizes as small as 10 nm,[18] which promises device applications with a lateral density exceeding Tera-bit/$in^2$, competitive with perovskite ferroelectrics, while the thickness scaling limit is the physical thickness of one monolayer.

We thank A. Gruverman, P. Dowben and A. Tagantsev for enlightening discussions, and A. Kuntz, H. Lu, and A. Rajapitamahuni for technical assistance. XH acknowledges the support from NSF Grant CAREER No. DMR-1148783 and Nebraska Research Council. SD acknowledges support from the DOE (#DE-SC0004530). XH and SD acknowledge the support from NSF MRSEC Grant No. DMR-0820521.

Figure 1

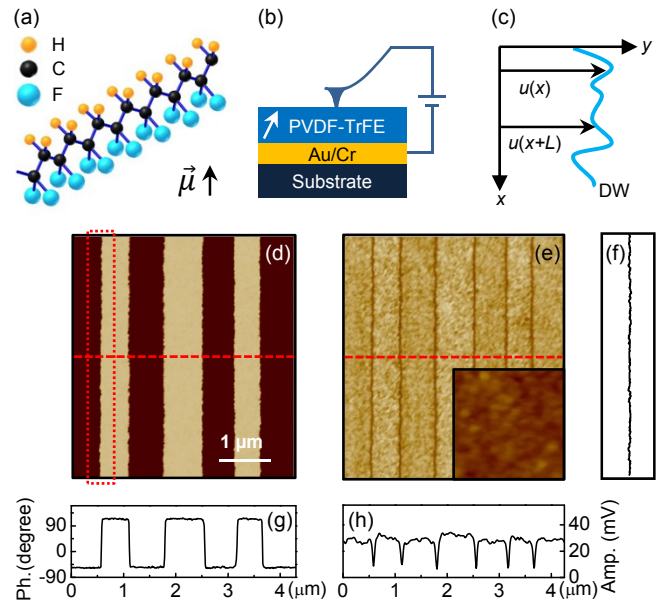

FIG. 1. (a) Schematic PVDF chain. (b) Schematic experimental setup. The arrow indicates the polarization direction. (c) Schematic of a rough DW. (d) Phase and (e) amplitude images of the same stripe-domains on a 10 ML PVDF-TrFE. The light (dark) domains are written with +10 V (-10 V). Inset: The topography image shows RMS roughness of ~9 Å. (f) DW extracted from the phase image shown in the red box in (d). (g) and (h) show the cross-sectional signal along the dashed lines in d) and e), respectively.

# Figure 2

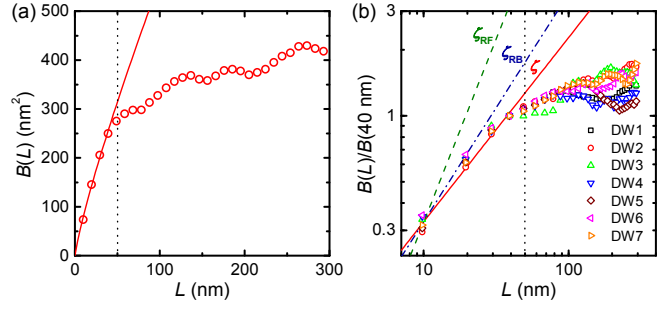

FIG. 2. (a) DW correlation function $B(L)$ vs. $L$ for a 15 ML PVDF-TrFE film. The solid line is a fit with $\sim L^{0.84}$. (b) Log-log plot of $B(L)$ normalized to $B(40$ nm$)$ vs. $L$ for 7 DWs on the 15 ML PVDF-TrFE film, and the fits using the average $\zeta$ (solid) and the values of $\zeta_{RB}$ (dash-dot line) and $\zeta_{RF}$ (dash line) predicated by Eqs. 1 for $d_{eff} = 1.5$. The dotted lines mark $L = 50$ nm.

Figure 3

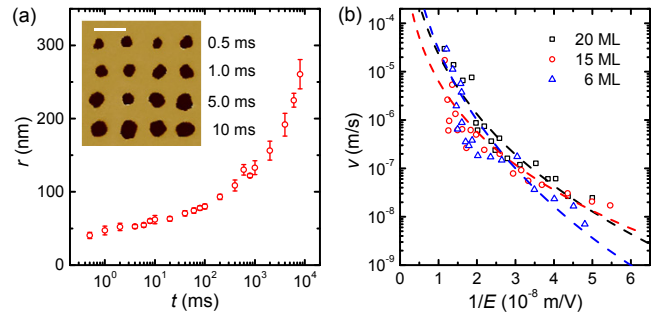

FIG. 3. (a) Dot-domain radius as a function of voltage pulse duration for a 15 ML PVDF-TrFE film. Inset: Dot-domain written with -11 V pulse with different durations. Scale bar: 200 nm. (b) DW velocity as a function of inverse electric field for the 20 ML, 15 ML, and 6 ML films, and the fits to the creep model (dashed lines).

Figure 4

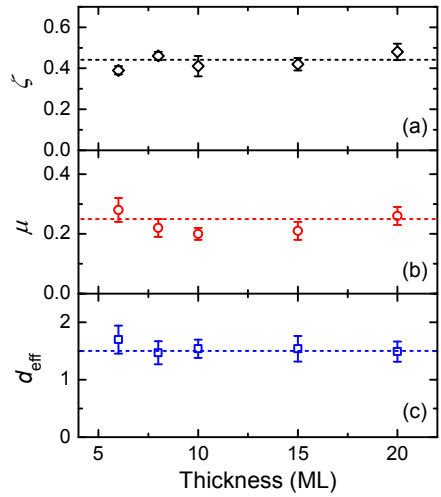

FIG. 4. (a) DW roughness exponent $\zeta$, (b) creep exponent $\mu$, and (c) effective dimensionality $d_{eff}$ of PVDF-TrFE films as a function of film thickness in the unit of ML. The dashed lines serve as the guide to the eye.

# Domain Wall Roughness and Creep in Nanoscale Crystalline Ferroelectric Polymer (Supplementary Information)

Z. Xiao[1], Shashi Poddar[1], Stephen Ducharme[1,2], and X. Hong[1,2*]

[1] Department of Physics and Astronomy, University of Nebraska-Lincoln, NE 68588-0299

[2] Nebraska Center for Materials and Nanoscience, University of Nebraska-Lincoln, NE 68588-0299

[*] xhong2@unl.edu

## Effect of the AFM Tip Size on the Domain Wall Imaging

The AFM cantilever used in this study has a nominal tip radius of ~20 nm. The actual contact area between the AFM tip and the polymer is much smaller. To find the critical length scale $L$ where the AFM tip size starts to interfere with the domain wall (DW) imaging, we performed the PFM experiments on a PVDF-TrFE film with sampling intervals of 5 nm. As shown in Fig. S1, it is clear that the data points above $L = 10$ nm follow well the power law dependence, while the size scaling breaks down for the $L = 5$ nm data point. This sharp softening of $L$-dependence in the correlation function $B(L)$ is consistent with the notion that if the imaging is limited by the contact size between the AFM tip and the polymer, the DW correlation function no longer depends on $L$. We thus conclude that the contact size is ~5 nm and only image the DWs using 10 nm sampling intervals to extract the roughness exponent $\zeta$.

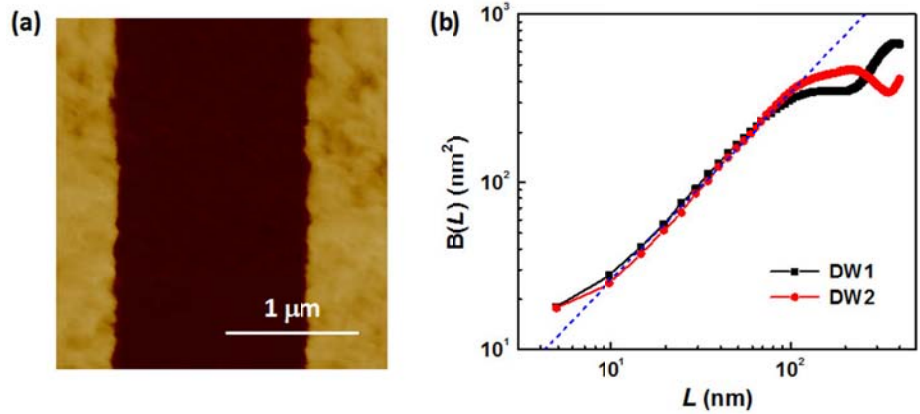

Fig. S1 (a) PFM image of DWs on a 10 monolayer PVDF-TrFE film with 5 nm sampling intervals. (b) The extracted correlation function $B(L)$ vs. $L$. The blue dashed line serves as the guide to the eye.

## Effect of the Cutoff Length on the Value of DW Roughness Exponent

The correlation function $B(L)$ of some DWs exhibits early softening below 50 nm, which gives rise to a roughness exponent that depends on the cutoff length for extracting the value of $\zeta$. In this case, the value of $\zeta$ extracted based on data between $L = 10 - 50$ nm only provides a lower bound of the roughness exponent. For example, for the data shown in Fig. 2(a) in the manuscript (Fig. S2(a)), we observe a clear increase in $\zeta$ value as we reduce the cutoff length from 50 nm to 30 nm (Fig. S2(b)). For other DWs, such as the one with $B(L)$ shown in Fig. S2(c), the value of $\zeta$ does not exhibit strong dependence on the cutoff length for $L$ below 50 nm (Fig. S2(d)).

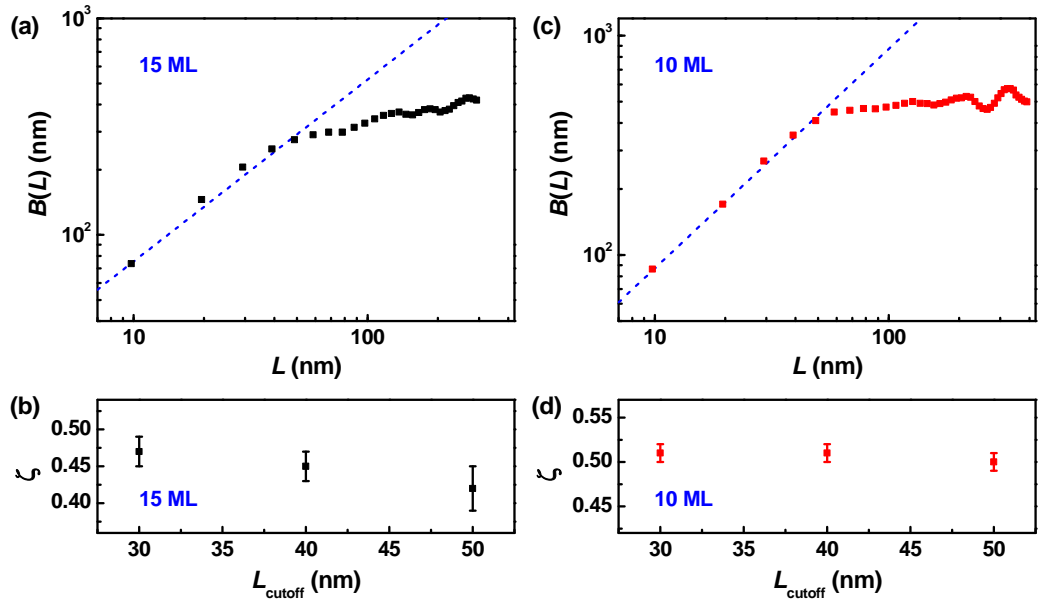

Figure S2 (a) DW correlation function $B(L)$ vs. $L$ shown in Fig. 2(a) for a DW on a 15 ML PVDF-TrFE film. (b) The extracted $\zeta$ value as a function of the cutoff length $L_{cutoff}$ from the $B(L)$ data in (a). (c) $B(L)$ vs. $L$ and (d) the corresponding $\zeta$ vs. $L_{cutoff}$ for a DW on a 10 ML film. The dashed lines are based on the $\zeta$ values extracted with $L_{cutoff}$ = 50 nm.